\RequirePackage{fix-cm}
\documentclass[smallextended]{svjour3}       % onecolumn (second format)
\smartqed  % flush right qed marks, e.g. at end of proof
\usepackage{graphicx}
 \journalname{ArXiv}
\begin{document}

\title{Quantitative assessment of drivers of recent climate variability: An information theoretic approach%\thanks{Grants or other notes
}
%\subtitle{Climate variability: Transfer Entropy}

\titlerunning{Climate variability: Transfer Entropy}        % if too long for running head

\author{Ankush Bhaskar \and  Durbha Sai Ramesh \and Geeta Vichare \and Triven Koganti \and S. Gurubaran
        %etc.
}


\institute{{Ankush Bhaskar \and  Durbha Sai Ramesh \and Geeta Vichare \and S. Gurubaran \at 
              Indian Institute of Geomagnetism, New Panvel, Navi Mumbai 410218, India. \\
              Tel.:  +91-22-2748 4000\\
              Fax: +91-22-27480762\\
              \email{ankushbhaskar@gmail.com} }    \\      %  \\
           \and
                      Triven Koganti \at
              Indian Institute of Technology, Kharagpur, India.
}

\date{Received: date / Accepted: date}

\maketitle

\begin{abstract}

Identification and quantification of possible drivers of recent climate variability remain a challenging task. This important issue is addressed adopting a non-parametric information theory technique, the \textit{Transfer Entropy} and its normalized variant.
It distinctly quantifies actual information exchanged along with the directional flow of information between any two variables with no bearing on their common history or inputs, unlike correlation, mutual information etc. Measurements of greenhouse gases,  $ CO_{2}$, $ CH_{4}$ and $N_{2}O$; volcanic aerosols; solar activity: \textit{UV} radiation, total solar irradiance (\textit{TSI}) and cosmic ray flux (\textit{CR});  El Ni\~{n}o Southern Oscillation (\textit{ENSO}) and Global Mean Temperature Anomaly (\textit{GMTA})  made during 1984-2005 are utilized to distinguish driving and responding climate signals. Estimates of their relative contributions reveal that $ CO_{2}$ ($\sim 24 \% $), $ CH_{4}$ ($\sim 19 \%$) and volcanic aerosols ($\sim23 \% $) are the primary contributors to the observed variations in \textit{GMTA}.  While, \textit{UV} ($\sim 9 \%$) and \textit{ENSO} ($\sim 12 \%$) act as secondary drivers of variations in the \textit{GMTA}, the remaining play a marginal role in the  observed recent climate variability. Interestingly,  \textit{ENSO} and \textit{GMTA} mutually drive each other at varied time lags. This study assists future modelling efforts in climate science.

\end{abstract}

\section{Introduction}
Global climate variability is of paramount importance to mankind and society as well. It is therefore widely discussed in public domain and is extensively studied by researchers. Several significant scientific investigations carried out during the past few decades attribute climate variability mainly to  atmospheric greenhouse gases, solar activity and aerosols from both natural and anthropogenic sources besides internal variability of the ocean-atmosphere system. Most of these studies report good correlation between variations in global mean temperature and climate variables such as greenhouse gases, cosmic rays, solar radiation, El Ni\~{n}o Southern Oscillation (\textit{ENSO}), global cloud cover, geomagnetic pole position, geomagnetic field, etc. \cite{herschel1801observations,dickinson1975solar,eddy1976maunder,timmermann1999increased,haigh1996impact,beer2000role,shindell2001solar,le2005long,usoskin2008cosmic,kerton2009climate,laken2012cosmic,robock2000volcanic,solomon2010persistence,hansen2000global,montzka2011non,dergachev2012impact}. These studies recognise four major drivers of the recent climate variability. One stems from the natural climate variables essentially connected to solar activity whereas the others are directly linked to anthropogenic sources such as the greenhouse gases and to volcanic aerosols and internal variability of the Ocean-Atmosphere system manifested as \textit{ENSO}, etc. Though the causal links between several of the above cited climate variables and variability in global mean temperature have been well studied, precise estimates of their quantitative influence still remain largely elusive and uncertain. It is important to realise that the cause-effect relationship inferred between the variables and climate variability are essentially based either on observed correlations or parametric modeling approaches. Techniques like linear correlation and mutual information which are symmetric in nature suffer from lack of information on the sense of directionality, while model-based methods are limited by model idealization \cite{verdes2005assessing}. Hence, it becomes rather difficult to decipher the true causal relationship between any two physical processes using these techniques. Therefore, many aspects related to climate variability and/or change still remain contentious and unresolved in the absence of reliable quantitative assessments of causality.

There are number of attempts to address the causality between natural and anthropogenic drivers using information theory approach \cite{kodra2011exploring,attanasio2012contribution,attanasio2012testing,das2012sea,stern2014anthropogenic,runge2014quantifying}. \cite{kodra2011exploring} showed that radiative forcing mainly due to $CO_2$ affects global temperature. The granger causality is applied to global climate data by \cite{stern2014anthropogenic}. They observed that both anthropogenic and natural forcing cause temperature change and as a feedback temperature changes affect greenhouse gas concentration.  Whereas, \cite{attanasio2012testing} showed that there is no signature of Granger causality from natural forcing to global temperature but Granger causality is observed from anthropogenic forcing to global temperature. So there is still ambiguity in understanding the role of the natural drivers in global temperature changes.

With this background, we investigate the relationship between various climate proxy records of both natural and anthropogenic origin adopting climatic time series analyses applying information theory-based stochastic methods which revolve around the concept of entropy or ensemble information content \cite{shannon1948mathematical}. In essence, the relationship between climate forcing variables related to solar activity, greenhouse gases and volcanic aerosols and the response signal i.e. Global Mean Temperature Anomaly (\textit{GMTA}) is explored and quantified in terms of the transfer of information with direction. Firstly, using \textit{transfer entropy} (\textit{TE}) with a directionality index \cite{schreiber2000measuring,de2011information}, we show that all the natural and anthropogenic proxies considered in this study indeed drive variations in the \textit{GMTA}. Further, their relative contributions in inducing the \textit{GMTA} are evaluated using normalized transfer entropy (\textit{NTE}) \cite{wang2011analyzing}. An important finding which emerges from this study is that volcanic aerosols (a natural source, henceforth referred as \textit{aerosols}) indeed compete with the well known greenhouse gases $CO_{2}$ and $CH_{4}$ in contributing to the \textit{GMTA} by accounting for a quarter ($\sim 25 \%$) of its variability. Among the remaining climate proxies studied here, \textit{ENSO} and \textit{UV} radiation together contribute $\sim 20 \%$ to the observed \textit{GMTA} with a similar share, while the rest are marginal.

%---------------------------------------------------------------------------------------------------------------------------------------------
\section{Data, Method and Analysis}
\subsection{Databases used}
In the present study, a total of nine proxies representing the recent climate variations for the epoch 1984-2005 are extracted from several databases and used. This time window was selected for two reasons. One, prominent climate changes are observed during this window and secondly, accurate simultaneous measurements of all these climate variables are available during this time period. The proxies used in this study are:  \textit{ENSO} index; global mean concentration of greenhouse gases, namely, $CO_{2}$, $CH_{4}$ and $N_{2}O$; global mean \textit{aerosol} optical depth at 550 nm; total solar irradiance (\textit{TSI}) along with solar ultraviolet flux (\textit{UV}) for wavelength window 120-400 nm; cosmic ray flux/neutron flux (\textit{CR}) and \textit{GMTA}. The global monthly mean data of greenhouse gases were obtained from the World Data Center for Greenhouse Gases (http://ds.data.jma.go.jp/gmd/wdcgg/wdcgg.html) whereas \textit{ENSO} index was collected from http://ncar.ucar.edu/. Global mean aerosol optical depth data were taken from http://data.giss.nasa.gov/modelforce/strataer/. \textit{TSI} and \textit{UV} fluxes were retrieved from the World Radiation Center (http://www.pmodwrc.ch/). The neutron flux is measured at Oulu Cosmic Ray Station (http://cosmicrays.oulu.fi/). The global mean temperature anomaly of the Earth was drawn from Met Office Hadley Center (http://www.metoffice.gov.uk/). All the proxies used are monthly mean values and are shown in Figure ~\ref{fig:Fig11}. These data show certain interesting features such as steady increasing trend in the \textit{GMTA} mimicking the greenhouse gases increase. The sudden increase in \textit{aerosol} optical depth seen in the figure is due to the atmospheric impact of the volcano eruptions (El Chichon, 1982, and Pinatubo, 1991) resulting in enhanced atmospheric opacity. The \textit{TSI}, \textit{UV} and cosmic rays clearly reflect the solar activity cycle. The \textit{TSI} and \textit{UV} peaks coincide with solar maxima while cosmic ray flux peaks correlate with the solar minimum. The \textit{ENSO} index shows occurrence of warm (positive) and cold (negative) phases across the Pacific associated with ocean currents during the studied epoch.

\subsection{Transfer Entropy and Directionality Index}

\cite{shannon1948mathematical} realized that the concepts of information and uncertainty are related to each other. He demonstrated that the occurrence of an event of lower probability (\textit{P}) indicates more information. This information can be characterized by the \textit{Information Entropy} or \textit{Shannon Entropy} and it is defined for a random variable $x$ as:
%\begin{linenomath*}
	\begin{equation}
	H_{x}=\sum\limits_{i=1}^{N} P_{i} (x) \log_{2} \bigg(\frac{1}{P_{i}(x)}\bigg)
	\end{equation}
%\end{linenomath*}
where  $ P_{i}(x)$ is  the probability of observing \textit{x} independently. Similarly, Shanon entropy, $H_{y}$  can be defined for the other variable $y$. To quantify actual information explicitly exchanged between these two variables $x$ and $y$, an information measure known as  \textit{Transfer Entropy} was  introduced by \cite{schreiber2000measuring}. This overcomes the limitations posed by measures of correlation and other entropy metrics by enabling us to distinctly quantify actual information exchanged along with the directional flow of information between any two variables with no bearing on their common history or inputs \cite{schreiber2000measuring,de2011information,das2012sea,balasis2013statistical,vichare2016equatorial}. The \textit{TE} from a process $x$ to another process $y$ after a time lag $\tau$ is the quantity of information that the state of $y$ has at a time $t + \tau$ based exclusively on the state of $x$ at time $t$. This can be represented by the following expression \cite{marschinski2002analysing}:\newline 
\textit{Transfer Entropy ($\tau$) =  \Big[ information about future observation of $y(t + \tau)$ gained from past joint observations of x and y\Big] -
	%\newline -
	\Big[information about future observation $y(t + \tau)$ gained from past observations of $y$ only \Big] = information flow from $x$ to $y$. } \\
Therefore, \textit{transfer entropy} between two random variables or processes represented by $x$ and $y$ is mathematically depicted as \cite{das2012sea}: 
%\begin{linenomath*}
	\begin{equation}
	TE_{x\rightarrow y}(\tau)=\sum\limits P (y(t+\tau),y(t),x(t)) \log_{2} \Bigg(\frac{P (y(t+\tau),y(t),x(t))* P(y(t))}{P (x(t),y(t))*P(y(t+\tau),y(t))}\Bigg)
	\end{equation}
%\end{linenomath*}
Similarly, \textit{transfer entropy} from $y$ to $x$,  $ TE_{y\rightarrow x} (\tau)$ can be estimated for different time lags/delays ($\tau$). 
Few interesting properties of \textit{TE} emerge from the above equation. These are listed as follows: Transfer entropy, \textit{TE} (a) is an asymmetric measure, (b) is based on transition probabilities (i.e. in a Markov process the probability of going from a given state to the next state.), hence incorporates the directionality of information flow, (c) is a measure of information transfer (exchange) rather than information shared, (d) enables quantifying information flow separately in both directions, and (e) is a model independent measure. Therefore, asymmetric nature of \textit{TE} can be used to detect the directed net flow of information between two physical processes represented as two time series. Depending on the magnitudes of $ TE_{x\rightarrow y} (\tau)$ and $ TE_{y\rightarrow x} (\tau)$ the driver and response signals/processes can thus be identified. The statistical significance of the obtained \textit{TE} is evaluated following the surrogate data test \cite{theiler1992testing,de2011information}. The null hypothesis to be tested is that there is no cause-effect relationship between the two variables (time series). This hypothesis is evaluated adopting a significance level of $5\%$ utilizing 100 randomized surrogate data-sets. 

In the presence of multiple drivers, estimation of relative contribution of each variable/proxy to the observed response signal assumes importance. To enable this, \cite{wang2011analyzing} have introduced a measure known as \textit{normalized transfer entropy} (\textit{NTE}) by accounting for the amount of information stored in $x(t)$ and $y(t+ \tau)$ which is given as \cite{das2012sea}: 
%\begin{linenomath*}
	\begin{equation}
	NTE_{x\rightarrow y}(\tau)= \frac{TE_{x\rightarrow y}(\tau)}{\sqrt{ H_{x}*H_{y+\tau}}}
	\end{equation}
%\end{linenomath*}
Similarly, $NTE_{y\rightarrow x}(\tau)$ can also be estimated. This \textit{NTE} enables comparison of contributions by several driver-response pairs. In the present research, the relative contribution to the global mean temperature anomaly (variable $y$) from climate proxies (variable $x$) are estimated using \textit{TE} and \textit{NTE}. Further, the directionality index defined by the following equation is used to compute the net flow of information, which is a convenient way to visualize the direction of net information flow between any two time series.
%\begin{linenomath*}
	\begin{equation}
	D_{x\rightarrow y}(\tau)= NTE_{x\rightarrow y}(\tau)-NTE_{y\rightarrow x}(\tau)
	\end{equation}
%\end{linenomath*}
The positive values of $D_{x\rightarrow y}$ indicate that the flow of information is in the direction of $x\rightarrow y$, suggestive of \textit{x} being a driver and \textit{y} being the response. Whereas, negative values of $D_{x\rightarrow y}$ would indicate \textit{y} as the driver and \textit{x} as the response signal.

\subsection{Analysis}

Prior to estimation of \textit{TE} and \textit{NTE}, the recorded non-stationary time series are interpolated for uniformity of data using cubic Hermite polynomial  to yield data with 10 samples per month. In order to minimize edge effects, 20 data points from both ends of the interpolated time series are discarded from the analysis. 
Following  \cite{carbone2004analysis,carbone2013information,carbone2007scaling}, these time series are transformed into stationary sequences. This is accomplished by subtracting the estimated moving averages from the original time series. The moving average is estimated using the formula:
%\begin{linenomath*}
	\begin{equation}
	\bar{x}_{n}(t)=\frac{1}{n} \sum\limits_{k=0}^n x(t-k)	
	\end{equation}
%\end{linenomath*}
Several trial moving average window sizes (n) were adopted for the analysis and a window size n=20 was found optimal to impart stationarity to the recorded time series used in this study. Since, moving avarge is low pass filter, the subtracted stationary time series now contains high frequency, the monthly stochastic variations of the variables. To estimate \textit{TE}, we need to arrive at the true probability distribution of a variable under consideration. We adopted the non-parametric histogram technique for this purpose. However, determination of an optimum bin-width becomes crucial in order to avoid over-smoothing arising from a large bin-width choice or empty bins because of too small a bin size.

An optimum bin-width is arrived using \cite{scott1979optimal} expression $w = 3.49 \sigma N^{-1/3}$, where $N$ is the number of data points and $\sigma$ is the standard deviation. Based on values of $w$ determined for each time series, the corresponding probability distributions are estimated. These are further used in the calculation of \textit{TE}. 

\section{Salient Results}

The \textit{transfer entropy} in both the directions corresponding to the global mean temperature anomaly and eight other proxies which represent natural and anthropogenic processes are computed. Figure ~\ref{fig:Fig33} shows the estimated \textit{TE} values for different time lags ($\tau$) related to various pairs along with their corresponding threshold of $5 \% $ statistical significance level. The calculated \textit{TE} values are observed unambiguously above the threshold significance. This suggests that the information flow is statistically significant. Therefore, we reject the null hypothesis and suggest that there exists a statistically significant cause-effect relationship between the pairs of proxies investigated in this study with \textit{GMTA} being always one of the time series. Further, note that the red curves representing the significative \textit{TE} from climate variable to \textit{GMTA} are generally higher than the blue lines which represent the transfer in the reverse flow direction. Clearly, for proxies $CO_{2}$, $CH_{4}$,  \textit{UV} and  \textit{aerosols} the red curves (\textit{TE} from proxy to \textit{GMTA}) are prominently above the blue curves (\textit{TE} from \textit{GMTA} to proxy). This  distinct segregation of \textit{TE} curves is not observed in the case of \textit{TSI}, cosmic rays and $ N_{2}O$. Interestingly, for \textit{ENSO}, the \textit{TE} in the direction \textit{GMTA} $\rightarrow$ \textit{ENSO} is higher after a time lag of $\sim 3$ months compared to that in the reverse direction. This suggests that both \textit{GMTA} and \textit{ENSO} perhaps interact variably at different time delays $\tau$. One common feature in all the plots is that \textit{TE}  drastically increases between delay times 0 and 1 month with steady values thereafter, indicating that the two variables affect almost instantaneously.

In summary, an important inference from above is that the flow of information related to the proxies; $CO_{2}$, $CH_{4}$, \textit{aerosols} and \textit{UV}; is significant and generally higher towards the \textit{GMTA} compared with the corresponding flow in the reverse directions. Thus, these four variables qualify to be the prominent drivers of the observed climate variability manifested as variations in \textit{GMTA}. In order to estimate the strength of net information flow for each driver, using  Equation 4, the directionality index ($D_{x\rightarrow y}$) is computed. The directionality index for each climate proxy as a function of time lag presented in Figure ~\ref{fig:Fig44} confirms that among climate variables $CO_{2}$, $CH_{4}$, \textit{aerosols} and \textit{UV}, the first three are indeed the primary drivers with competing strength while \textit{UV} is a relatively weak driver. It is surprising to note that \textit{aerosols} which were hitherto not so well studied do contribute to \textit{GMTA} as much as the greenhouses gases, $CO_{2}$ and $CH_{4}$. 

To estimate percent contribution of each climate proxy to the observed variability in the global mean temperature anomaly, we computed the integrated net information flow (directionality index using Equation 4) employing time lags of 3 months and 10 months which yielded consistent results. Here, $100\%$ sum is calculated by considering the information flow of studied drivers alone. Moreover, we do not rule out the possibility of other variables which could influence the present climate variability. Therefore, the estimated percentage of each driver is relative in nature. Note that $100\%$ sum does not mean that the complete variance of \textit{GMTA} is explained by studied drivers, it is the total normalized information flow to \textit{GMTA}. 
%Using these values, we arrived at percent contribution of each driver signal (i.e. proxies) to the \textit{GMTA}. 
The pie chart shown in Figure ~\ref{fig:Fig55} corresponds to percent contributions calculated for a lag of 3 months and clearly demonstrates the hierarchy of contribution by each variable to the observed global climate variability. It can be clearly seen that greenhouse gases ($CO_{2}$ and $CH_{4}$) and \textit{aerosols} are the chief contributors to the observed recent climate variability with near similar ($\sim 19-24 \% $) percent contributions. While, \textit{UV} and \textit{ENSO} which together explain nearly $\sim 21\% $ variations in \textit{GMTA} qualify as secondary contributors. The remaining candidate climate variables \textit{CR}, $N_{2}O$ and \textit{TSI} are marginal players in the observed climate variability. These interesting results are discussed in the following section in the context of our current understanding of climate variability.
\section{Discussion and Conclusions}
The two principal issues addressed here are: (1) identification of primary drivers  of recent (1984-2005) climate variability (\textit{GMTA}) among the eight climate variables considered in this study and  (2) quantification of their influence on the climate variability. Transfer entropy and its variant are used to address these issues. Use of this technique in climate studies is a recent trend  \cite{verdes2005assessing,knuth2013revealing,das2012sea,runge2012quantifying}. Its application to study the recent climate variability in particular is in its infancy. For example, \cite{verdes2005assessing} limiting their study to greenhouse gases and \textit{TSI} conclude that the former are the significant contributors to the present climate change, while \cite{das2012sea} showed that during cooler climate epochs of the interglacial Marine Isotope Stages (MIS), the sea surface temperature was more similar to the atmospheric $CO_{2}$ forcing. Recently, \cite{runge2014quantifying} proposed the method based on the graphical model to identify the causal connection, strength of interaction and the time delays. In the present work, the recent climate variability is investigated in a more inclusive manner by incorporating \textit{aerosols}, inter-variability of the atmosphere-ocean system and several manifestations of solar activity as proxies and estimating their relative contributions.

\begin{figure}
	\begin{center}
		\includegraphics[height=20pc,angle=0]{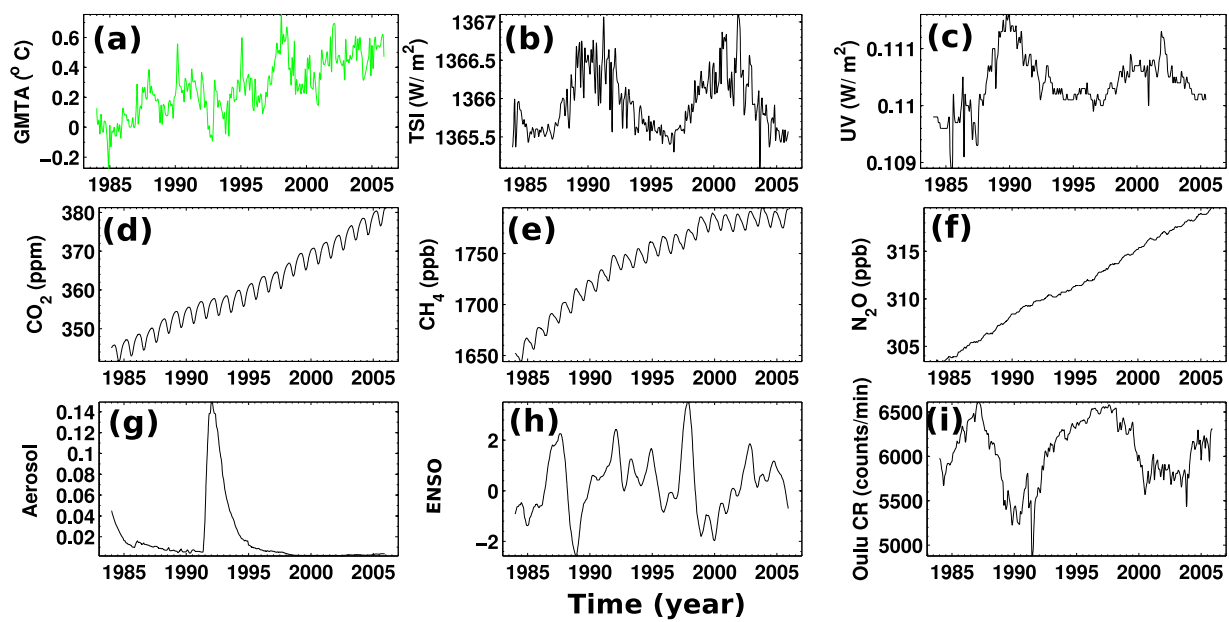} 
		\caption{ Time-series of various climate proxies employed in this study for year 1984-2005: (a) Global mean temperature anomaly (\textit{GMTA}), (b) Total solar irradiance (\textit{TSI}), (c) Ultraviolet irradiance (\textit{UV}), concentration of (d) Carbon dioxide ($ CO_{2}$), (e) Methane ($CH_{4}$) and (f) Nitrous oxide ($N_{2}O $) (g) Stratospheric aerosol optical depth, (h) \textit{ENSO} index and (i) Cosmic ray flux (\textit{CR}) at Oulu neutron monitor station. %The rapid increase in global concentration of greenhouse gases and solar activity dependence of cosmic ray flux and solar radiation is very much evident. 
			The cause-effect relationships between \textit{GMTA} (shown in green) and the remaining climate variables (shown in black) are investigated.}
		\label{fig:Fig11} 
	\end{center}
\end{figure}

Our study quantifies the effect of greenhouse gases ($ CO_{2}$ + $ CH_{4}$) on recent climate variability with $\sim 43\%$ contribution. Interestingly, \textit{aerosols} which alone contribute $\sim 23 \%$ to the observed climate signal qualify as the other primary driver in addition to $ CO_{2}$ and $ CH_{4}$. The internal climate forcing by \textit{ENSO} with $\sim12\%$ contribution and \textit{UV} share of $\sim 9\%$ owing to solar variability constitute the minor components of climate variability that deserve attention. These relative contributions are consistent with earlier reports by \cite{mende1994solar,hofmann2006role,verdes2007global}. Evidently, the other components of solar variability, \textit{TSI} and \textit{CR} together with the greenhouse gas $ N_{2}O$ with contributions $\sim 5\%$ are the marginal components. It is pertinent to note, though the greenhouse gas $ N_{2}O$  has stronger radiative forcing per molecule and long residence time ($\sim 120$ years), its atmospheric concentration is less (see Figure ~\ref{fig:Fig11} ). This could be the reason for $ N_{2}O$  being among the weakest contributors. In the following, we briefly discuss possible mechanisms by which \textit{aerosols}, \textit{ENSO}, and \textit{UV} influence the climate variability represented by $\textit{GMTA}$. 

Atmospheric \textit{aerosols} as one of the competing prime drivers of \textit{GMTA} is brought out by the data used in this study, since the data had contributions from at least two volcanic episodes. Atmospheric \textit{aerosols} originate from volcanic eruptions and have relatively less concentration compared to anthropogenic aerosols. However, as they are released at higher altitudes resulting in their longer residence time in the atmosphere \cite{robock2000volcanic}, they can affect the climate variability on different time scales. The initial effect of \textit{aerosols} is net cooling near the surface, as they stay in the atmosphere for few years having e-folding time of $\sim 1-2$ years. These \textit{aerosols} also enhance the destruction of ozone, and cause stratospheric warming which could contribute to possible global climate variability. It is therefore generally difficult to comprehend their exact atmospheric impact and feedback mechanisms owing to several complexities \cite{robock2000volcanic} through model based studies. Hence we are content with quantifying the percent contribution of \textit{aerosols}  to recent climate variability, which forms an important input to future modeling studies.

\begin{figure}
	\begin{center}
		\includegraphics[width=28pc]{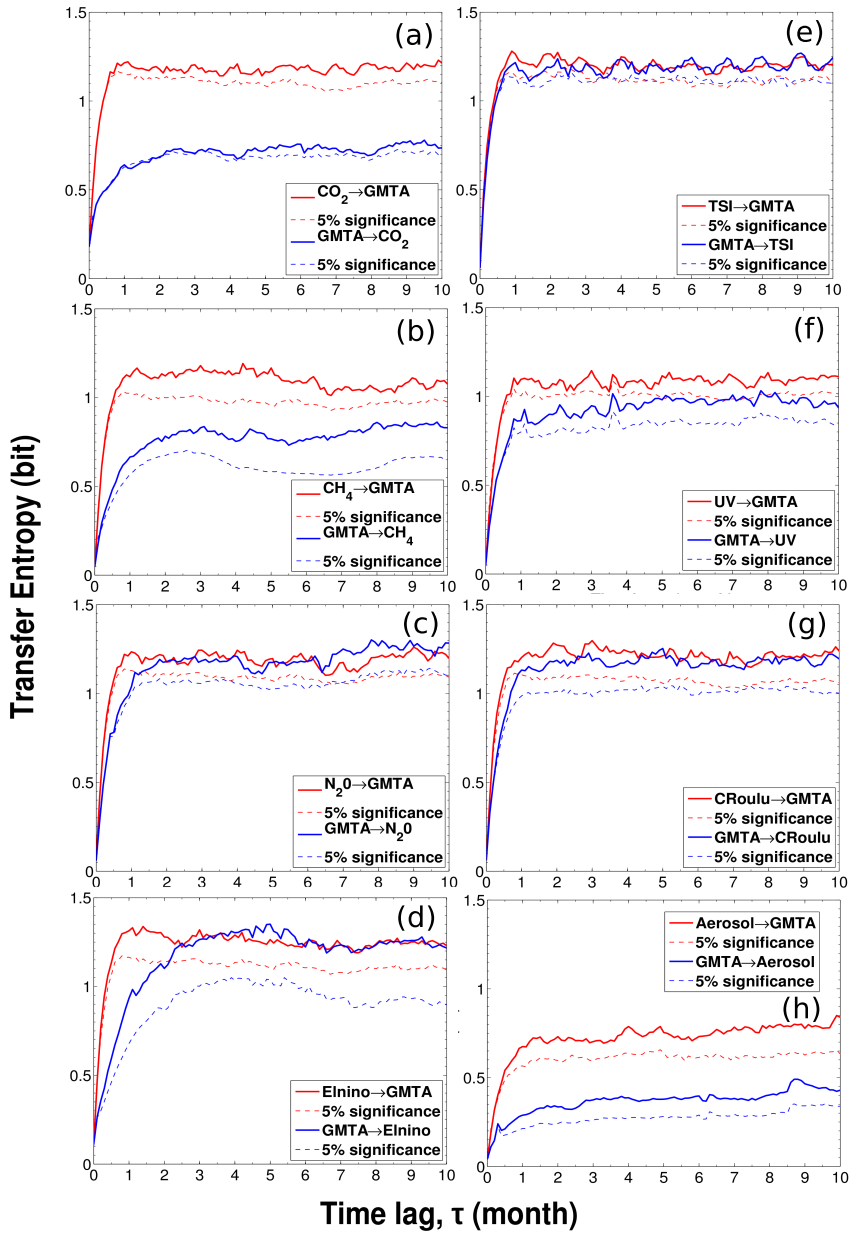} 
		\caption{ \textit{Transfer entropy (TE)} between various climate proxies and \textit{GMTA}. The red lines represent the information flow from the studied climate variables to \textit{GMTA} whereas blue line is representation of information flow from \textit{GMTA} to the climate variables. The dashed line is the $5\%$ significance level, constructed from 100 surrogate datasets. %All \textit{TE} values are statistically significant. 
			Note that the net flow of information from remaining proxies to \textit{GMTA} is generally higher than that in the reverse direction.
		}
		\label{fig:Fig33} 
	\end{center}
\end{figure}

\begin{figure}
	\begin{center}
		\noindent\includegraphics[width=40pc]{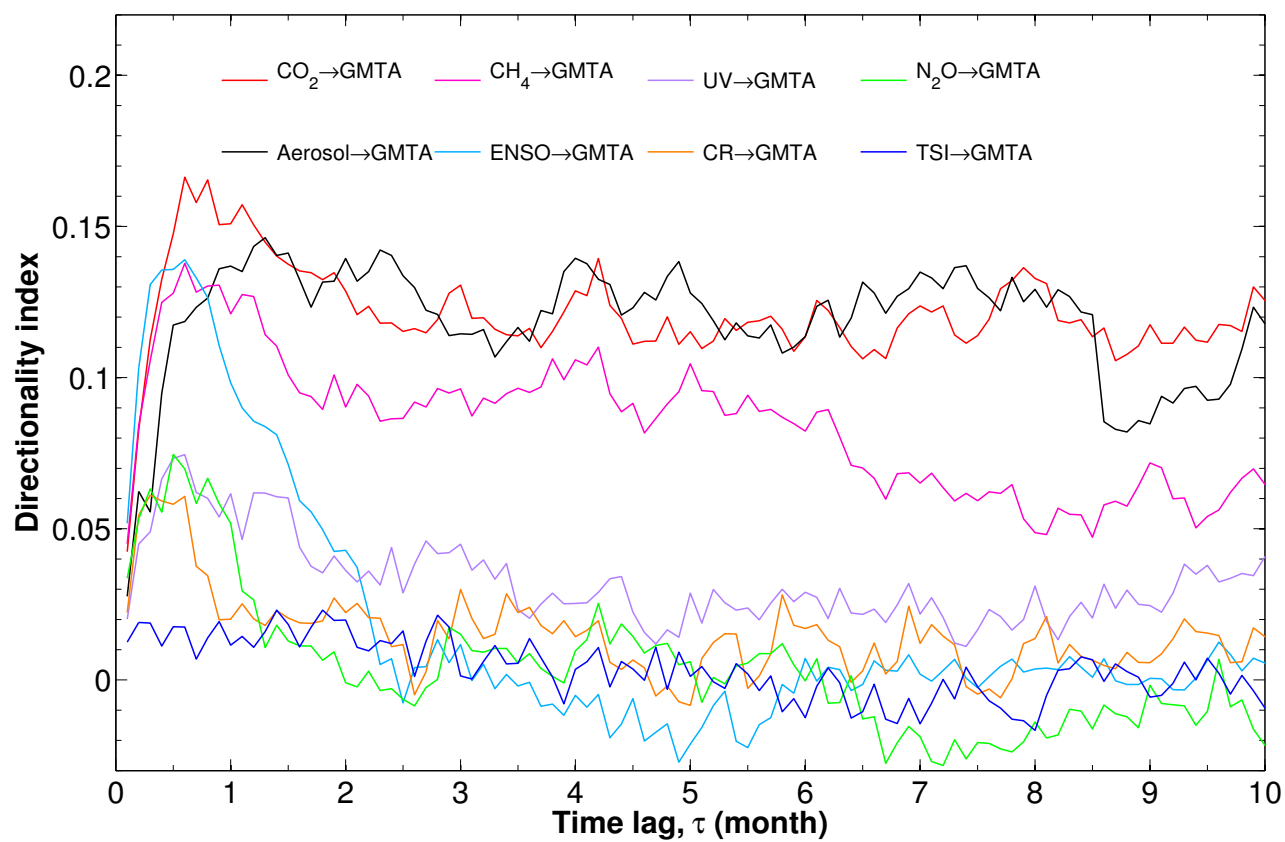} 
		\caption{ Directionality Index ($D_{x\rightarrow y}$) of information transfer from the climate proxies to the \textit{GMTA} as a function of time lag, $\tau$. The magnitude and sign of the index indicate the strength of information transfer and its direction respectively. %Here, positive values indicate that the net flow of information is from studied variable to the \textit{GMTA}. 
			Note that at different time lags the net information flow from $CO_{2}$, $CH_{4}$ and aerosol to \textit{GMTA} is dominant. 
		}
		\label{fig:Fig44} 
	\end{center}
\end{figure}

\begin{figure}
	\begin{center}
		\includegraphics[width=40pc]{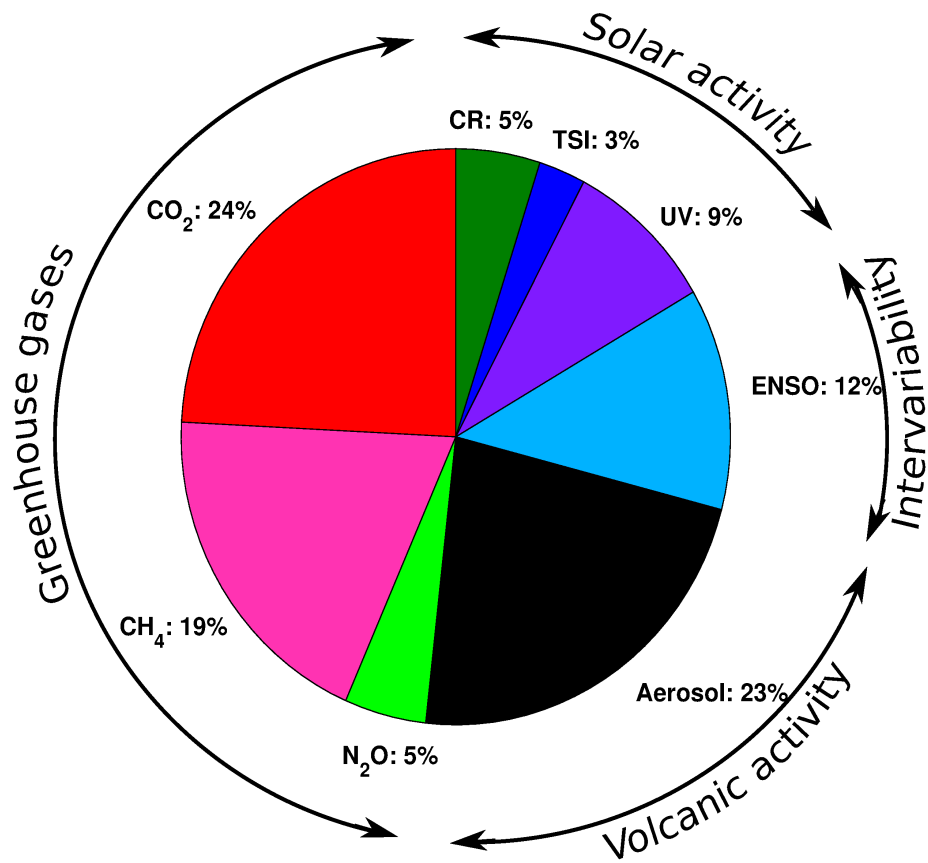} 
		\caption{Percent contribution of various drivers estimated using integrated directionality index for 0-3 month time lag. The dominance of greenhouse gases ($CO_{2}$ and  $CH_{4}$) and volcanic aerosols is evident.} %The greenhouse gases alone account for $\sim 50 \%$ of the recent observed climate variability whereas remaining variability is due to volcanic aerosols, solar activity and inter-variability of the ocean-atmosphere system. Solar \textit{UV} radiation seen to be more effective in affecting \textit{GMTA} compare to \textit{TSI}. The individual  contributions of total solar irradiance (\textit{TSI}), cosmic ray flux (\textit{CR}) and \textit{$N_{2}O$} to the recent climate variability is marginal.
		\label{fig:Fig55} 
	\end{center}
\end{figure}

Among the secondary drivers of \textit{GMTA}, \textit{ENSO} is more prominent than \textit{UV}. Identification of \textit{ENSO} as a driver or response signal remained contentious with conflicting results \cite{cobb2003nino,timmermann1999increased}. The occurrences of \textit{ENSO} during the last millennium are explained invoking  natural variability of the ocean-atmosphere system \cite{cobb2003nino}, while  frequent occurrences of \textit{ENSO} in last few decades and in a climate model are attributed to the greenhouse warming \cite{trenberth1997nino,timmermann1999increased,cai2014increasing}. One may say that the effect on global climate due to ENSO is positive at some places and negative in another places. However,it has been reported that ENSO and super ENSO periods cause increase in global temperature \cite{lean2009will,lean2010cycles}.
Interestingly, our \textit{transfer entropy} estimates shown in Figure ~\ref{fig:Fig33}d, indicate the cross-talk between \textit{ENSO} and \textit{GMTA} at different time lags is mixed in nature. That said, at lags till 3 months \textit{ENSO} drives \textit{GMTA}, while, at later time lags \textit{GMTA} assumes the role of a driver. Thus, \textit{ENSO} and \textit{GMTA} mutually affect each other differently at varied time lags by exchanging their roles. While the mechanism of \textit{ENSO} driving \textit{GMTA} is better understood, it is unclear how \textit{GMTA} drives occurrence of \textit{ENSO}. We speculate that variations in \textit{GMTA} perturb the energy (heat) budget of the ocean-atmosphere system that would affect the occurrence of \textit{ENSO}.

Based on percent contributions derived from our analysis, the solar \textit{UV} flux influences \textit{GMTA} more than \textit{TSI}. This is in consonance with simulation studies reported by \cite{haigh1996impact}. During a solar cycle, the amplitude of variation in total solar irradiance (\textit{TSI}) is on the order $<0.1\%$, while it attains the highest amplitude in the ultraviolet (\textit{UV}) radiation band reaching $32\%$ \cite{lean1989contribution,haigh1996impact,beer2000role}. Therefore, solar radiation in the \textit{UV} band can affect the Earth's climate distinctly more than TSI. It is well established that the solar \textit{UV} radiation affects the atmosphere through changes in the photochemical dissociation rate and ozone heating. This is clearly brought out in the simulation studies by \cite{haigh1996impact} adopting a general circulation model. However, the nature of coupling between the stratosphere and the troposphere remains a topic of investigation. In this context, it is important to note that the current phase of decreasing solar cycle amplitude should give rise to a global cooling effect. However, as observed in this study and by others (e.g. \cite{mende1994solar,hofmann2006role,verdes2007global}), the solar radiative forcing is small compared to that of greenhouse gases resulting in the dominance of the effects of greenhouse gases on climate variability. Moreover, greenhouse gases like $ CO_{2}, N_{2}O$ etc have large residual life times in the atmosphere. In particular, $CO_{2} $ can reside in the atmosphere for 100-1000 years \cite{montzka2011non}. This would lead to warming as greenhouse gases accumulate in the atmosphere and counter the cooling effect due to the low solar flux. Even, the initial short duration cooling effects due to \textit{aerosols}, which are short lived, may not be adequate to mask the warming induced by enhancement in greenhouse gases of longer life.

The solar activity modulates the incident flux of cosmic rays on the Earth's atmosphere through changes in the interplanetary magnetic field. Cosmic rays could affect the global cloud cover through ionization resulting in changes in global temperature \cite{carslaw2002cosmic,tinsley2008global}. Present study reveals that cosmic rays contribute $\sim 5\% $ to \textit{GMTA} variability. Albeit small, these quantitative estimates obtained for the period of 1984-2005 point that cosmic rays do contribute to \textit{GMTA}. Also, it supports the earlier studies \cite{dickinson1975solar,carslaw2002cosmic,tinsley2008global} which report influence of cosmic rays on the global temperature.

In summary, the present study unambiguously establishes that greenhouse gases and \textit{aerosols} contribute dominantly to the global mean temperature anomaly. The greenhouse gases alone account for $\sim 50 \%$ of the recent observed climate variability. The forcing due to the Sun in \textit{UV} band and inter variability of the atmosphere-ocean system (\textit{ENSO}) plays a secondary role in climate variability. As an interesting sidelight, all the constituents of natural forcings together seem to make contributions equal to the anthropogenic sources (greenhouse gases) in the context of recent climate variability. In spite of the present low solar activity, our results imply that global warming would continue mainly due to increase in greenhouse gases forcing and decline in volcanic aerosols. Due to the well documented increase in greenhouse gases in the atmosphere and  mutual driving ability of \textit{GMTA} and \textit{ENSO} as revealed in this study, frequent future occurrence of \textit{ENSO} remains a distinct possibility independent of its internal variability. The application of transfer entropy to climate proxies can be further used to identify other drivers of climate variability and understand their relative importance.

\begin{acknowledgements}
Authors thank  World Data Center of Greenhouse Gases (http://ds.data.jma.go.jp/gmd/wdcgg/wdcgg.html), World Radiation Center (http://www.pmodwrc.ch/), Oulu Cosmic Ray Station (http://cosmicrays.oulu.fi/), National Geophysical Data Center, Met office, Hadley Center, UK (http://www.metoffice.gov.uk/) and Goddard Space Flight Center Sciences and Exploration Directorate Earth Sciences Division (http://data.giss.nasa.gov/modelforce/strataer/) for making necessary data available in public domain. Authors gratefully acknowledge Joanna Haigh, Imperial College, London for valuable discussions and constructive comments on the manuscript.
\end{acknowledgements}

\newpage

% \begin{figure}
% \begin{center}
%  \noindent\includegraphics[width=40pc]{Fig2.eps} 
%   \caption{ The neutron flux (\textit{CR}),total IMF (B) and solar wind velocity (Vsw) are shown for November,2001 FD event. 
%  The total interplanetary magnetic field (B) and solar wind speed (Vsw) are plotted with inverted sign for better 
%   comparison with neutron flux. The decrement seen in neutron flux is accompanied with sharp variations in B and Vsw. 
%   The first vertical line from left represents arrival of interplanetary shock or onset, second vertical line represents first minimum of 
%   respective quantities and third vertical line indicates recovery times for respective quantities.}
%   \label{fig:Fig22} 
%  \end{center}
%  \end{figure}

\clearpage

\bibliographystyle{spmpsci} %abbrvnat acm
\bibliography{entropy}

\end{document}